\documentclass[%
 reprint,
 amsmath,amssymb,
 aps,prl
]{revtex4-1}

\usepackage{graphicx}% Include figure files
\usepackage{dcolumn}% Align table columns on decimal point
\usepackage{bm}% bold math
\usepackage{physics}
\usepackage{mleftright}
\usepackage{siunitx}
\usepackage{makecell}

\usepackage{epstopdf}

\usepackage{ulem}
\usepackage[usenames]{xcolor}

\definecolor{mygreen}{RGB}{0,204,102}

\DeclareMathAlphabet{\mathpzc}{OT1}{pzc}{m}{it}

\begin{document}

\preprint{APS/123-QED}

\title{Antiferromagnetic Kitaev Interaction in \textit{f}-Electron Based Honeycomb Magnets}

\author{Seong-Hoon Jang}
\author{Ryoya Sano}
\author{Yasuyuki Kato}
\author{Yukitoshi Motome}
\affiliation{
 Department of Applied Physics, The University of Tokyo, Tokyo 113-8656, Japan
}

\date{\today}

\begin{abstract}
We theoretically propose a family of $f$-electron based magnets that realizes Kitaev-type bond-dependent anisotropic interactions. Based on {\it ab initio} calculations, we show that $A_2$PrO$_3$ ($A$: alkali metals) crystalize in a triclinic structure with honeycomb layers of edge-sharing PrO$_6$ octahedra. Each Pr$^{4+}$ cation has a $4f$ electron in the $\Gamma_7$ doublet, which comprises a spin-orbital entangled Kramers pair with the effective moment $J_{\rm eff}=1/2$. By using the Wannier orbitals from the {\it ab initio} calculations, we find that the effective interactions between the $J_{\rm eff}=1/2$ moments are predominantly of {\it antiferromagnetic} Kitaev type for light alkali metals $A$=Li and Na, in stark contrast to the ferromagnetic ones in $4d$- and $5d$-electron based materials. Our finding would provide a playground for the Kitaev spin liquids that is hard to be accessed by the candidates ever discovered.
%\begin{description}
%\item[Usage]
%Secondary publications and information retrieval purposes.
%\item[PACS numbers]
%May be entered using the \verb+\pacs{#1}+ command.
%\item[Structure]
%You may use the \texttt{description} environment to structure your abstract;
%use the optional argument of the \verb+\item+ command to give the category of each item. 
%\end{description}
\end{abstract}

\pacs{Valid PACS appear here}% PACS, the Physics and Astronomy
                             % Classification Scheme.
%\keywords{Suggested keywords}%Use showkeys class option if keyword
                              %display desired
\maketitle

%\tableofcontents

%\section{\label{sec:level1}Introduction}

Quantum spin liquid (QSL) is an exotic magnetic phase in a system with interacting localized spins~\cite{AN1973, BA2010, ZH2017A, SA2017}. In the QSL, spins remain disordered even at zero temperature because of competing interactions and strong quantum fluctuations, while they are strongly correlated and quantum entangled. The quantum entanglement gives rise to a topological order~\cite{WE1991A, LE2006} and fractionalization of spins into emergent quasiparticles~\cite{SA1992, NA2008}. The peculiar nature of the QSLs has attracted growing interest, since it is potentially utilized in quantum computation~\cite{KI2003}.

The QSL has been long sought for antiferromagnets with triangular-based lattice structures, where the antiferromagnetic interactions compete with each other due to geometrical frustration~\cite{LA2011, DI2013}. Several candidates have been intensively studied, e.g., in triangular~\cite{SH2003, KU2005}, kagome~\cite{HI2001, SH2005, OK2009}, hyperkagome~\cite{OK2007}, and pyrochlore compounds~\cite{MO2007, RO2011}. Another route to the QSLs has also been pursued for magnets with directionally dependent interactions~\cite{KU1982, KH2005}. Such directional dependence originates from the coupling between the spin and orbital degrees of freedom, which may lead to severe competition even for nonfrustrated lattice structures~\cite{KH2003, NU2015, OL2015}.

Recently, a model in the latter category, called the Kitaev model, has attracted upsurge interest, as it provides an exact QSL ground state~\cite{KI2006B}. The model has directionally dependent Ising interactions on the three types of bonds in a honeycomb structure. Such peculiar interactions can arise in the presence of the strong spin-orbit coupling when two requisites are fulfilled~\cite{JA2009}: (i) localized electrons in spin-orbital entangled states and (ii) quantum interference between the indirect hopping processes of the electrons via ligands. For instance, these are approximately realized in some $4d$- and $5d$-electron based materials with the $d^5$ low-spin configuration, such as $A_2$IrO$_3$ ($A$=Na, Li)~\cite{SI2010, SI2012} and $\alpha$-RuCl$_3$~\cite{PL2014, KU2015}. In these materials, (i) is realized by the $J_{\rm eff}=1/2$ Kramers doublet under the octahedral crystal field, and (ii) is by two different $d$-$p$-$d$ paths in the edge-sharing honeycomb network of the ligand octahedra.

While the study of the Kitaev QSLs has been rapidly growing, the candidate materials are still limited. The two requisites above are not exclusive to the $d^5$ low-spin systems. In fact, several efforts were recently made to explore another candidates. For instance, the $d^7$ high-spin configuration was proposed to support the Kitaev-type interactions~\cite{LI2018, SA2018}. Also, $f$-electron systems, where the strong spin-orbit coupling is expected, were nominated as potential candidates~\cite{LI2017, RA2018}. However, material-oriented researches along these directions are not fully carried on. It is highly desired to design another platform for the Kitaev QSL for accelerating the cooperative studies between experiments and theories.

In this Letter, we theoretically propose a family of Kitaev candidates in $f$-electron based oxides. For satisfying the requisite (i), we consider an $f^1$ configuration under the octahedral crystal field, which results in the spin-orbital entangled $\Gamma_7$ doublet. For (ii), we postulate such $f^1$ materials with the edge-sharing honeycomb network similar to the $4d$ and $5d$ candidates. We substantiate these ideas by {\it ab initio} calculations for the rare-earth oxides $A_2$PrO$_3$ ($A$: alkali metals). By deriving the low-energy effective model, we find that the magnetic property is dominated by the Kitaev-type interactions for light alkali metals $A$=Li and Na, and remarkably, the interactions are antiferromagnetic, in stark contrast to the ferromagnetic ones presumed for the $4d$ and $5d$ candidates~\cite{JA2009, YA2014, WI2016}. The qualitative difference originates from the peculiar spatial anisotropy of the $f$ orbitals as well as the weak crystal field. Our results not only add the Kitaev candidates but also provide the possibility of antiferromagnetic Kitaev QSLs, which have recently attracted much attention owing to the intriguing properties in an applied magnetic field~\cite{ZH2017B, GO2018, NA2018, RO2018, HI2018}.

\begin{figure}[t]
\includegraphics[width=0.85\columnwidth]{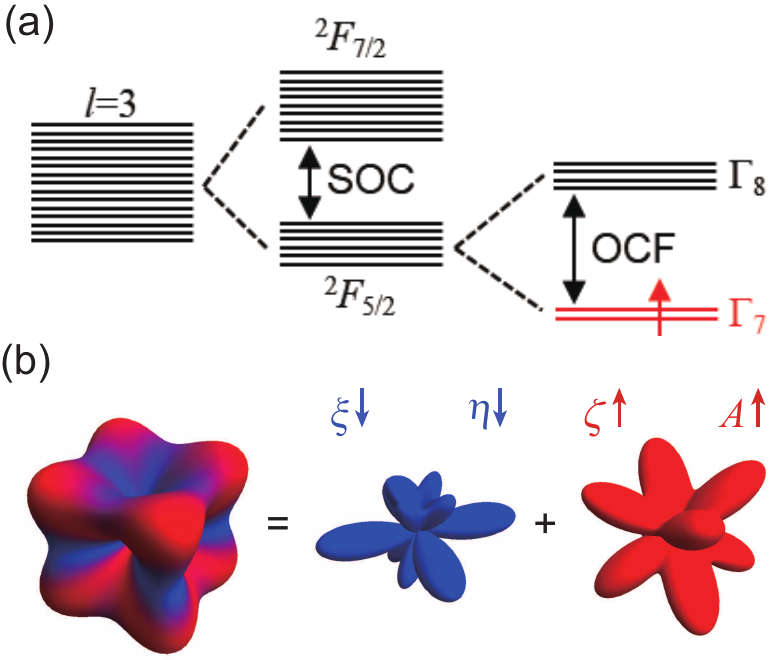}
\caption{\label{fig:f1}(a) $f^{1}$ level splitting by the spin-orbit coupling (SOC) and the octahedral crystal field (OCF). (b) Density profile of an electron in the pseudospin up state $\ket{+}$ for the $\Gamma_{7}$ doublet; see Eq.~(\ref{eq:+}).}
\end{figure} 

Let us first discuss the electronic state of an $f$ electron subject to the spin-orbit coupling and the octahedral crystal field. The 14 energy levels of the $f$ orbitals with the angular momentum $l=3$ are split into the low-energy $^{2}F_{5/2}$ sextet and the high-energy $^2{F}_{7/2}$ octet by the spin-orbit coupling [see the left and middle panels of Fig.~\ref{fig:f1}(a)]. The $^{2}F_{5/2}$ sextet is further split into the low-energy $\Gamma_{7}$ doublet and the high-energy $\Gamma_{8}$ quartet under the octahedral crystal field [see the right panel of Fig.~\ref{fig:f1}(a)]. Hence, the ground state for the $f^1$ configuration is given by the $\Gamma_7$ Kramers doublet.

The $\Gamma_{7}$ doublet can be treated as pseudospins with the effective moment of $1/2$, similar to the $J_{\rm eff}=1/2$ states in the $d^5$ low-spin case. The time-reversal pair is represented by 
\begin{align}
&\ket{+}=\frac{1}{\sqrt{21}}(2ic^{\dagger}_{\xi\downarrow}-2c^{\dagger}_{\eta\downarrow}+2ic^{\dagger}_{\zeta\uparrow}+3c^{\dagger}_{A\uparrow})
\ket{0}
,
\label{eq:+}\\
&\ket{-}=\frac{1}{\sqrt{21}}(2ic^{\dagger}_{\xi\uparrow}+2c^{\dagger}_{\eta\uparrow}-2ic^{\dagger}_{\zeta\downarrow}+3c^{\dagger}_{A\downarrow})
\ket{0},
\label{eq:-}
\end{align}
where {($\it{\xi}$, $\it{\eta}$, $\it{\zeta}$) and $A$ denote the $f$ orbitals with the irreducible representations T$_{2u}$ and A$_{2u}$, respectively~\cite{[We follow the conventional notations for the $f$ orbitals${\rm ,}$ e.g.${\rm ,}$ used in ]TA1980}, and $c_{\nu\sigma}^\dagger$ is a creation operator of an electron with orbital $\nu$ and spin $\sigma$. Figure~\ref{fig:f1}(b) displays the pseudospin state $\ket{+}$, which has a similar profile to the $d^5$ $J_{\rm eff}=1/2$ state~\cite{JA2009} but different directional spin dependence. Then, the pseudospin operator $\mathbf{S} = (S^x,S^y,S^z)^{\textrm{T}}$ can be defined by
\begin{equation}
S^\mu=-\frac{3}{5}
\begin{bmatrix} 
\mel{+}{J^\mu}{+} & \mel{+}{J^\mu}{-} \\
\mel{-}{J^\mu}{+}  & \mel{-}{J^\mu}{-}  
\end{bmatrix}
=\frac{1}{2}\sigma^\mu,
\label{eq:pseudospin}
\end{equation}
where $\mathbf{J}$ and $\boldsymbol{\sigma}$ are the total angular momentum operator and the Pauli matrix, respectively.

The above observation leads us to consider analogous Kitaev systems to the $4d$ and $5d$ materials by using the $f^1$ $\Gamma_7$ Kramers doublet. As an isostructural candidate with the iridium oxides $A_2$IrO$_3$, we consider Pr-based oxides $A_2$PrO$_3$ ($A$: alkali metals) with the $4f^1$ electron configuration in the Pr$^{4+}$ cations. We investigate the stability of the Pr oxides by {\it ab initio} calculations, and check if the electronic states realize the $\Gamma_7$ doublet. The \textit{ab initio} calculations with the structure optimization are performed by using \texttt{Quantum ESPRESSO}~\cite{GI2017}, and the maximally-localized Wannier functions (MLWFs) are extracted with \texttt{WANNIER90}~\cite{MO2014}. See Supplemental Material for further details~\cite{DM0000}. We calculated the compounds with $A$=Li, Na, K, Rb, and Cs, and found that all the results converge onto a triclinic structure with P\={1} symmetry. In the following, we focus on two materials with $A$=Li and Na since they are most interesting from the viewpoint of the effective magnetic couplings as discussed later. The comprehensive analyses including other compounds will be reported elsewhere. 

\begin{figure}[t]
\includegraphics[width=0.85\columnwidth]{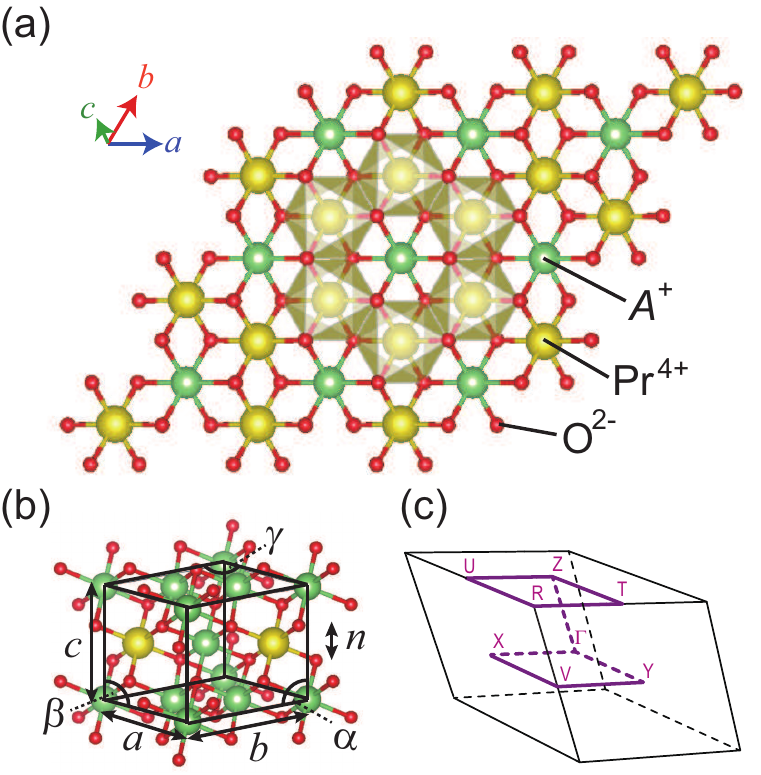}
\caption{\label{fig:f2}(a) and (b) The optimized triclinic structure for $A_2$PrO$_3$ with $A$=Li. The green, yellow, and red spheres denote $A^{+}$, Pr$^{4+}$, and O$^{2-}$ ions, respectively. The edge-sharing network of PrO$_6$ octahedra is partially shown. In (b), the black lines represent a primitive unit cell with the lattice parameters; $n$ is the average distance of the O layers sandwiching the Pr layer. (c) The first Brillouin zone for the triclinic structure. The purple lines represent the symmetric lines used in Fig.~\ref{fig:f3}.}
\end{figure} 

\begin{table}[t]
\caption{\label{tab:table1}Structural parameters of the optimized structures for $A_2$PrO$_3$ ($A$=Li, Na). See Fig.~\ref{fig:f2}(b) for the definitions of $a$, $b$, $c$, $\alpha$, $\beta$, $\gamma$, and $n$. The ratio $a/n$ becomes $3/\sqrt{2}\simeq 2.12$ in an ideal edge-sharing octahedra under the O$_h$ symmetry. $d_{\textrm{Pr-Pr}}$ and $\it{\theta}_{\textrm{Pr-O-Pr}}$ denote the averages of the Pr-Pr bond length and the Pr-O-Pr bond angle, respectively, for the neighboring Pr pair within the same honeycomb layer.}
\begin{ruledtabular}
\begin{tabular}{ccc}
&Li$_2$PrO$_3$&Na$_2$PrO$_3$\\
\hline
$a$ (\si{\angstrom})&5.6228&5.9950\\
$b$ (\si{\angstrom})&5.6270&5.9967\\
$c$ (\si{\angstrom})&5.1487&5.9923\\
$\it{\alpha}$ (\si{\degree})&79.701&80.199\\
$\it{\beta}$ (\si{\degree})&100.29&99.802\\
$\it{\gamma}$ (\si{\degree})&59.981&60.015\\
\hline
$n$ (\si{\angstrom})&2.3656&2.3328\\
$a/n$&2.3769&2.5698\\
\hline
$d_{\textrm{Pr-Pr}}$ (\si{\angstrom})&3.2473&3.4628\\
$\it{\theta}_{\textrm{Pr-O-Pr}}$ (\si{\degree})&95.175&102.15\\
\end{tabular}
\end{ruledtabular}
\end{table}

The optimized lattice structures of $A_2$PrO$_3$ ($A$=Li, Na) are composed of honeycomb layers of edge-sharing PrO$_6$ octahedra, as exemplified in Fig.~\ref{fig:f2}. The structural parameters are summarized in Table~\ref{tab:table1}. Each Pr layer is close to a perfect honeycomb structure with C$_3$ symmetry. We note that there are small trigonal distortions, indicated by the deviations of $a/n$ and $\theta_{\textrm{Pr-O-Pr}}$ from their ideal values $3/\sqrt{2}$ and $90$\si{\degree}, respectively. 

\begin{figure}[t]
\includegraphics[width=1.0\columnwidth]{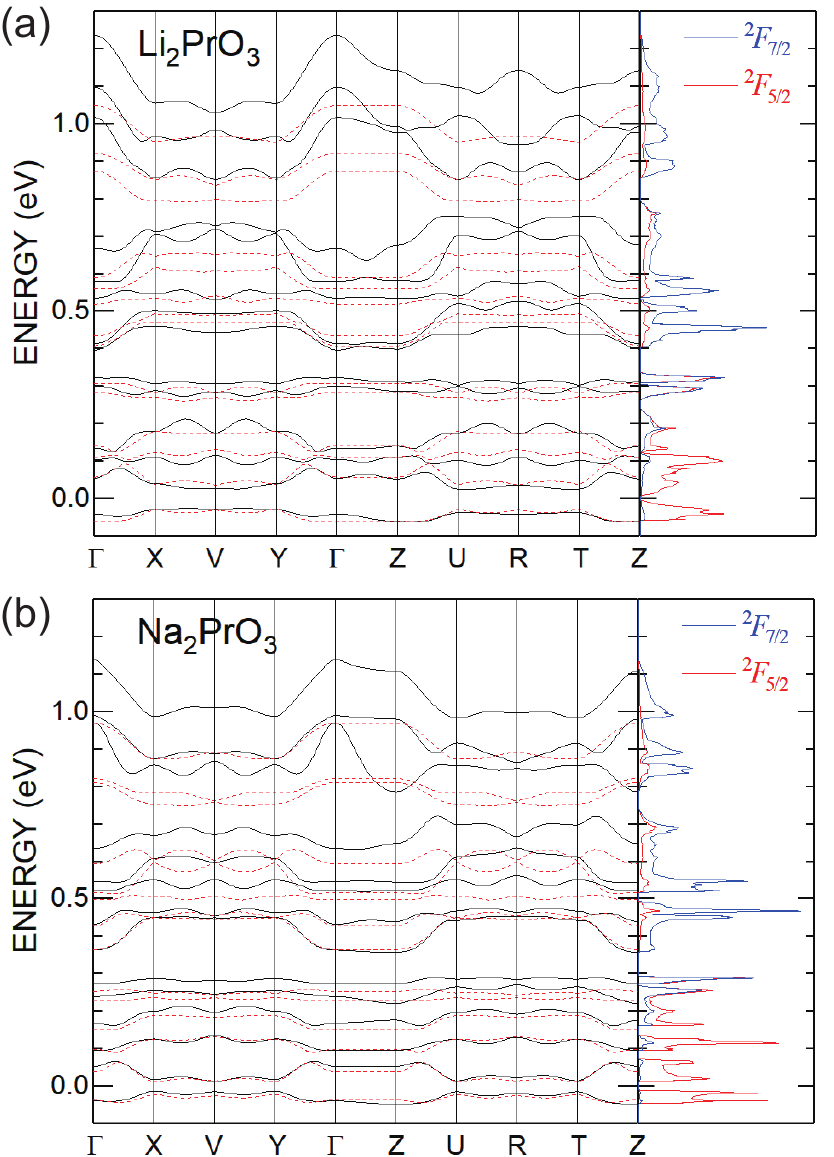}
\caption{\label{fig:f3}Electronic band structures for (a) Li$_2$PrO$_3$ and (b) Na$_2$PrO$_3$. The black solid and red dashed lines show the band dispersions obtained by the {\it ab initio} calculation and the tight-binding calculation with nearest-neighbor transfers estimated by the MLWFs, respectively. The Fermi level is set to zero. The right panels display the projected density of states to the $^2F_{5/2}$ and $^2F_{7/2}$ manifolds in the Pr $4f$ states.}  
\end{figure} 

The electronic band structures obtained by the {\it ab initio} calculations are shown in Fig.~\ref{fig:f3}. In both $A$=Li and Na cases, the Pr $4f$ bands are well isolated from the lower-energy O $2p$ bands and the higher-energy Pr $6s$, $A$ $2p$, and $A$ $2s$ bands~\cite{DM0000}. The bandwidth is slightly wider in the Li case, reflecting the smaller lattice constants in Table~\ref{tab:table1}. As expected in Fig.~\ref{fig:f1}(a), the energy bands originating from the $^2F_{5/2}$ sextet and the $^2F_{7/2}$ octet are split by the strong spin-orbit coupling; see the projected density of states in the right panels of Fig.~\ref{fig:f3}. In comparison with the results by nonrelativistic calculations, the spin-orbit coupling coefficient is estimated as $\sim120$~meV, close to the empirical values~\cite{HI1994, PO1996}. In the $4f^1$ state, the lowest-energy shallow bands (doubly degenerate) separated from the others lie below the Fermi level in both Li and Na cases. Thus, the results indicate that the systems are band insulators with two $f$ electrons per unit cell on average. We confirm that the MLWFs for the occupied states have the $\Gamma_7$-like profile.

In Fig.~\ref{fig:f3}, we also show the tight-binding band structures with transfer integrals between neighboring Pr cations estimated from the MLWFs in Fig.~\ref{fig:f3}~\cite{DM0000}. The {\it ab initio} results are well reproduced, especially for the low-energy bands, indicating that further-neighbor hoppings are less important because of} the localized nature of $4f$ orbitals.

The above analysis suggests that the $4f^1$ compounds may become the spin-orbit coupled Mott insulators under strong electron correlations. We here consider an effective model for the $\Gamma_7$ pseudospins in Eqs.~(\ref{eq:+}) and (\ref{eq:-}) by the second-order perturbation in terms of the nearest-neighbor transfer integrals between Pr cations in the same honeycomb layer~\cite{DM0000}. The calculations are performed by taking into account all the 91 intermediate $4f^2$ states of Pr$^{3+}$ whose multiplet levels are treated by the Russel-Saunders scheme following the literature~\cite{FR1962}. In the perturbation calculation, we take into account both the direct $4f$-$4f$ and the indirect $4f$-$2p$-$4f$ paths; we symmetrize the transfer integrals for three different directions by taking their average so as to recover the C$_3$ symmetry that is weakly broken in our optimized structures.

\begin{figure}[t]
\includegraphics[width=0.85\columnwidth]{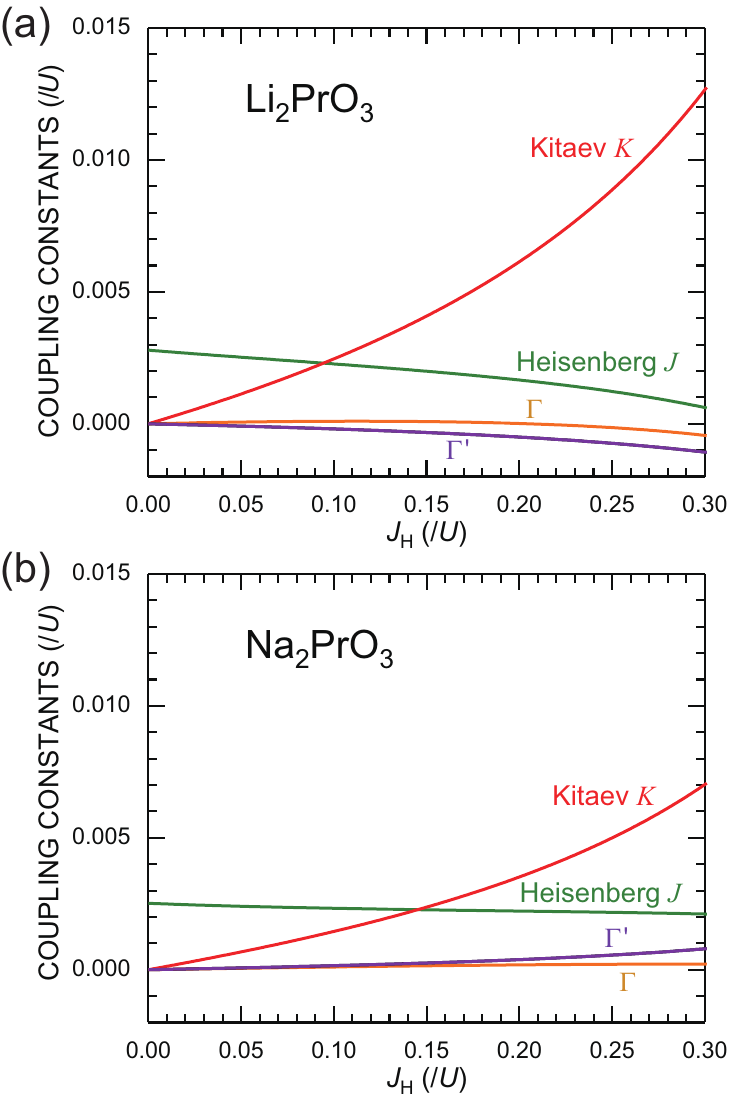}
\caption{\label{fig:f4}The coupling constants in the effective pseudospin Hamiltonian in Eq.~(\ref{eq:Heff}) for (a) Li$_2$PrO$_3$ and (b) Na$_2$PrO$_3$ as functions of the Hund's-rule coupling $J_{\rm H}$. The green, red, orange, and purple lines represent the Heisenberg $J$, Kitaev $K$, and off-diagonal couplings $\Gamma$ and $\Gamma^{\prime}$, respectively. All the energy scales are normalized by the intraorbital Coulomb repulsion $U$.}
\end{figure}

The effective pseudospin Hamiltonian for one of three types of bonds on the honeycomb structure ($z$ bond) is given in the matrix form
\begin{equation}
\mathpzc{H}^{ \left( z \right) }_{i,j}=
\mathbf{S}_i^{\rm T}
\begin{bmatrix} 
\it{J} & \Gamma & \Gamma^{\prime} \\
\Gamma & \it{J} & \Gamma^{\prime} \\
\Gamma^{\prime}  & \Gamma^{\prime} & \it{J}+\it{K}
\end{bmatrix}
\mathbf{S}_j,
\label{eq:Heff}
\end{equation}
where $\mathbf{S}_i$ is the pseudospin defined in Eq.~(\ref{eq:pseudospin}) at site $i$. The total Hamiltonian is given by the sum over the neighboring $\mu=x,y,z$ bonds, $\mathpzc{H}_{\rm eff} = \sum_{\mu} \sum_{{\langle i,j \rangle}_{\mu}} \mathpzc{H}^{(\mu)}_{i,j}$, where $\mathpzc{H}^{(x)}_{i,j}$ and $\mathpzc{H}^{(y)}_{i,j}$ are given by cyclic permutations of $\{ xyz \}$ in $\mathpzc{H}^{(z)}_{i,j}$. In Eq.~(\ref{eq:Heff}), the diagonal terms $J$ and $K$ represent the isotropic Heisenberg coupling and the bond-dependent anisotropic Kitaev coupling, respectively, and $\Gamma$ and $\Gamma^\prime$ are the symmetric off-diagonal couplings~\cite{YA2014}. The coupling constants are plotted in Fig.~\ref{fig:f4} as functions of the ratio of the Hund's-rule coupling $J_{\rm H}$ to the intraorbital Coulomb repulsion $U$. We find that, as increasing $J_{\rm H}/U$, the Kitaev coupling $K$ is largely enhanced, while $J$ is slightly reduced, and both $\Gamma$ and $\Gamma^\prime$ are small: the Kitaev interaction is most dominant in the large $J_{\rm H}/U$ region. Remarkably, $K$ is always positive, namely, {\it antiferromagnetic}. This is in stark contrast to the $d^5$ materials in which $K$ is considered to be ferromagnetic~\cite{JA2009, YA2014, WI2016}. Hence, the effective pseudospin model for the $4f^1$ systems is approximately given by the Kitaev-Heisenberg model with the antiferromagnetic $K$.

\begin{figure}[t!]
\includegraphics[width=1.\columnwidth]{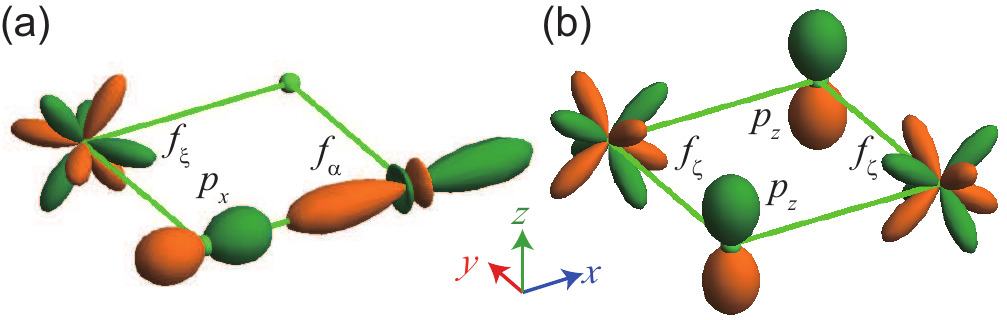}
\caption{\label{fig:f5}Relevant indirect hopping processes along a $z$ bond: (a) between $f_\xi$ and $f_\alpha$ orbitals via $p_x$ and (b) between $f_\zeta$ orbitals via $p_z$.} 
\end{figure} 

By carefully examining the perturbation processes, we find that two types of indirect hopping paths predominantly contribute to the antiferromagnetic Kitaev coupling: for a $z$ bond, one is $f_\xi$-$p_x$-$f_\alpha$ (equivalent to $f_\eta$-$p_y$-$f_\beta$) [Fig.~\ref{fig:f5}(a)] and the other is $f_\zeta$-$p_z$-$f_\zeta$ [Fig.~\ref{fig:f5}(b)]~\cite{[We follow the conventional notations for the $f$ orbitals${\rm ,}$ e.g.${\rm ,}$ used in ]TA1980}. 
The former $f_\xi$-$p_x$-$f_\alpha$ looks similar to the indirect $t_{2g}$-$p$-$e_g$ hopping processes ($d_{xy}$-$p_x$-$d_{3x^2-r^2}$) in the $d^5$ low-spin case~\cite{CH2013,SA2018}. The $t_{2g}$-$p$-$e_g$ processes contribute to the antiferromagnetic $K$, but the contribution is usually small because of the large crystal field splitting between the $t_{2g}$ and $e_g$ manifolds, typically larger than $1$~eV. In contrast, $f_\xi$-$p_x$-$f_\alpha$ in the present case can largely contribute to the antiferromagnetic $K$, as the crystal field splitting between $f_\xi$ and $f_\alpha$ is small $\sim 0.1$~eV, as shown in Fig.~\ref{fig:f3}. Meanwhile, the latter $f_\zeta$-$p_z$-$f_\zeta$ apparently resembles $d_{yz}$-$p_z$-$d_{zx}$ in the $d^5$ case which brings about the ferromagnetic $K$~\cite{JA2009}. However, the distinct spatial anisotropy of the $f_\zeta$ orbital allows the indirect hopping between the same orbitals, and contribute differently from the $d^5$ case. Thus, the level scheme and the spatial anisotorpy of the $f$ orbitals play a crucial role in the peculiar antiferromagnetic Kitaev coupling. 

Our finding of the dominant antiferromagnetic $K$ would be important, since the known candidates for the Kitaev QSL are presumed to be ferromagnetic. Although other interactions may stabilize a parasitic long-range order, the pristine effect of the anitiferromagnetic $K$ could be revealed, e.g., by thermal fluctuations~\cite{NA2015} and an applied magnetic field. The latter has recently attracted much attention owing to the possibility of a field-induced state, which does not appear for the ferromagnetic $K$~\cite{ZH2017B, GO2018, NA2018, RO2018, HI2018}.  

Finally, let us comment on the material trend in $A_2$PrO$_3$. As inferred by the comparison between Li and Na, the Kitaev coupling $K$ becomes smaller for $A$ with the larger ionic radius. This is mainly because the Pr-Pr distance $d_{\textrm{Pr-Pr}}$ becomes longer. At the same time, the trigonal distortion becomes larger ($a/n$ and $\theta_{\textrm{Pr-O-Pr}}$ become larger), which leads to relatively large other couplings $J$, $\Gamma$, and $\Gamma^\prime$. Hence, the $A$=Li case is optimal for the dominant antiferromagnetic Kitaev coupling within this series of compounds. The comprehensive analyses will be reported elsewhere. 

To summarize, we have proposed a class of the $f$-electron based Kitaev-type honeycomb magnets on the basis of the {\it ab initio} calculations and the effective model analysis. We found that $A_2$PrO$_3$ ($A$: alkali metal) is well described by the Kitaev-Heisenberg model in the low-energy sector of the $\Gamma_7$ Kramers doublet for the $4f^1$ configuration of Pr$^{4+}$ cations. We showed that the peculiar spatial anisotropy of the $f$ orbitals and the weak crystal field make the Kitaev coupling antiferromagnetic, in sharp contrast to the ferromagnetic ones in the $4d$ and $5d$ Kitaev candidates ever discussed. Our results provides a platform for the Kitaev QSL, which enables to access the parameter space beyond the existing candidates. 

\begin{acknowledgments}
The authors thank T. Miyake and H. Shinaoka for fruitful discussion. Parts of the computation have been done using the facilities of the Supercomputer Center, the Institute for the Solid State Physics, the University of Tokyo. The crystal structures in Figs.~\ref{fig:f2}(a) and \ref{fig:f2}(b) are visualized by \texttt{VESTA}~\cite{MO2011}. The second-order perturbation calculations are performed by using \texttt{SNEG} package~\cite{ZI2011}. This work is supported by Grant-in-Aid for Scientific Research under Grant No.~16H02206.
\end{acknowledgments}

% The \nocite command causes all entries in a bibliograWphy to be printed out
% whether or not they are actually referenced in the text. This is appropriate
% for the sample file to show the different styles of references, but authors
% most likely will not want to use it.
%\nocite{*}

\bibliography{Antiferromagnetic_Kitaev_f-electron}% Produces the bibliography via BibTeX.

\clearpage
\onecolumngrid

\appendix
\vspace{15pt}
\begin{center}
{\large \bf ---Supplemental Material---}
\end{center}

\setcounter{figure}{0}
\setcounter{equation}{0}
\setcounter{table}{0}
\renewcommand{\thefigure}{S\arabic{figure}}
\renewcommand{\theequation}{S\arabic{equation}}
\renewcommand{\thetable}{S\Roman{table}}
\baselineskip=6mm

\section{Details of {\it ab initio} calculations}

In the \textit{ab initio} calculations, we adopt the pseudopotentials of scalar-relativistic norm-conserving von Barth-Car type~\cite{BA1985}, non-relativistic norm-conserving Hartwigesen-Goedecker-Hutter type~\cite{HA1998}, and full-relativistic ultrasoft projector-augmented-wave-method Perdew-Zunger type~\cite{PE1981, BL1994} for $A$(=Li, Na), O, and Pr, respectively. We set the kinetic energy cutoff at 250~Ry. The lattice structures are optimized starting from the structural parameters for Rb$_2$CeO$_3$ listed in \texttt{Materials Project}~\cite{JA2013}. In the structure optimization, the criteria for the maximum crystal stress is set at 0.1~GPa. The remnant maximum atomic forces are less than 0.009~Ry/Bohr in the {\it ab}-plane and less than 0.0003~Ry/Bohr along the axis perpendicular to the plane. In the self-consistent field calculations for the structure optimization and the non-self-consistent field calculations for the electronic band structures, the (projected) density of states, and the construction of MLWFs, we use the Monkhorst-Pack grids~\cite{MO1976} of 4${\times}$4${\times}$4 and 8${\times}$8${\times}$8 ${\bf k}$-points, respectively. The convergence threshold in the self-consistent field calculations is set at 1.0${\times}10^{-10}$~Ry.

\section{Electronic band structures}

We show the electronic band structures and the projected density of states for Li$_2$PrO$_3$ and Na$_2$PrO$_3$ in Figs.~\ref{fig:fs1} and \ref{fig:fs2}, respectively. In each case, (a) displays the results in a wide energy range from $-25$~eV to $20$~eV, and (b) is for the middle energy range from $-6$~eV to $2$~eV, including the Pr $4f$ bands and the O $2p$ bands. Figure~3 in the main text shows the narrow energy range from $-0.1$~eV to $1.3$~eV, focusing on the Pr $4f$ bands hybridized with the O $2p$ bands. In both compounds, the $4f$-$2p$ bands are isolated from other bands, which facilitates the MLWF construction and the effective model analysis in the main text. 

In Li$_2$PrO$_3$ in Fig.~\ref{fig:fs1}, the hybridized bands of  Pr $6s$, Li $2s$, and Li $2p$ orbitals are located above $3.0$~eV, and the bands in the range of $3.1\sim4.0$~eV are mostly ascribed to Li $2p$ orbitals. The Pr $4f$ bands hybridized with the O $2p$ bands lie in the range of $-0.1\sim1.3$~eV, where the localized nature of the $f$ orbitals is manifested in the narrow bandwidth and the relatively high projected density of states. The main O $2p$ bands are located in the range of $-5.7\sim-2.0$~eV with weak hybridization with the Pr $4f$ bands. The bands in the deep energy levels of $-21.1\sim-13.9$~eV are, mainly from the hybridization of Pr $5p$ and O $2s$.

The overall feature of the band structure is shared with Na$_2$PrO$_3$, as shown in Fig.~\ref{fig:fs2}. In Na$_2$PrO$_3$, the hybridized bands of Pr $6s$ and Na $3s$ orbitals are located above $3.6$~eV, and the bands lie in the range of $1.7\sim3.6$~eV are mostly ascribed to Na $3s$ orbitals. The Pr $4f$ bands hybridized with the O $2p$ bands in the range of $-0.1\sim1.2$~eV are well separated from the main O $2p$ bands in the lower range of $-5.2\sim-2.2$~eV. The bands in the deep energy levels of $-21.1\sim-14.0$~eV are mainly from the hybridization of Pr $5p$ and O $2s$.

\section{Transfer integrals estimated from MLWFs}

\begin{figure}[%t
b]
\includegraphics[width=0.7\columnwidth]{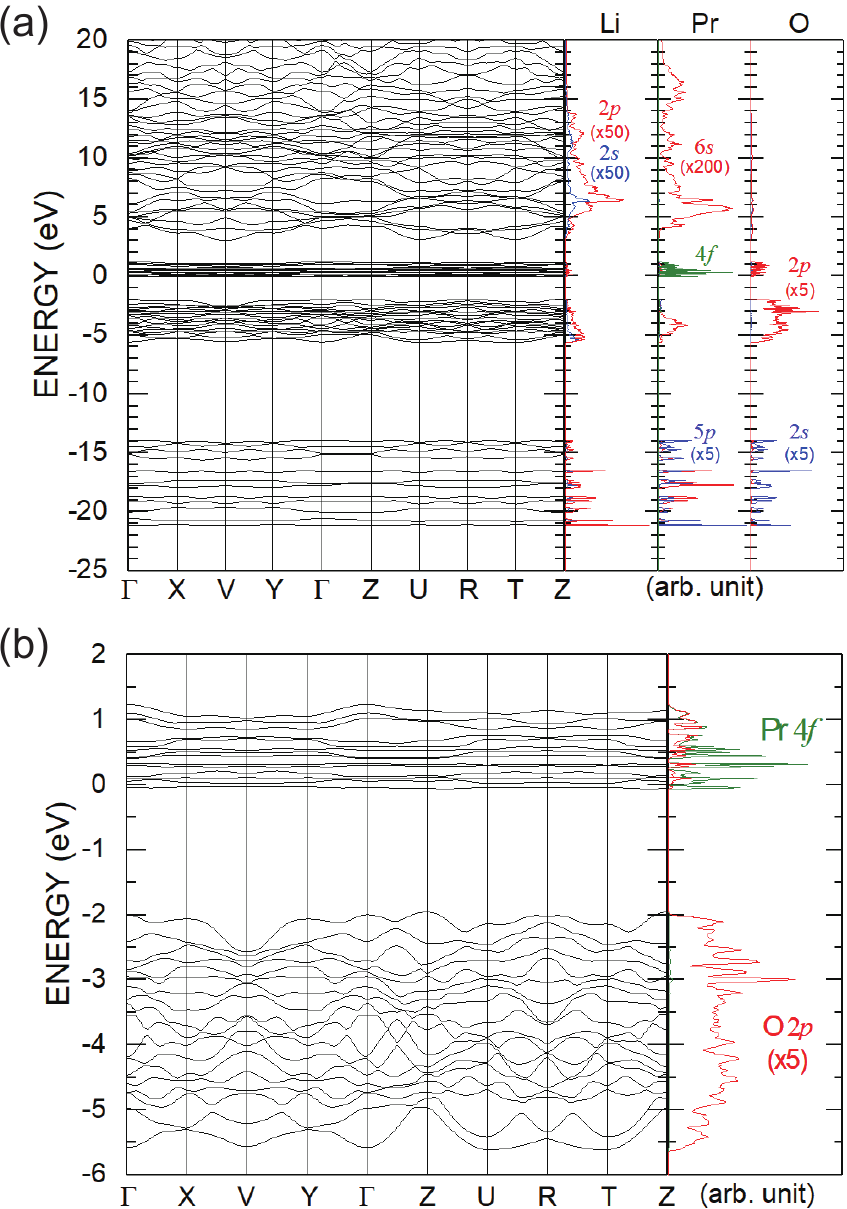}
\caption{\label{fig:fs1}Electronic band structures and the projected density of states for Li$_2$PrO$_3$ in the energy range from (a) $-25$~eV to $20$~eV and (b) $-6$~eV to $2$~eV. The Fermi level is set to zero.}
\end{figure}

\begin{figure}[t]
\includegraphics[width=0.7\columnwidth]{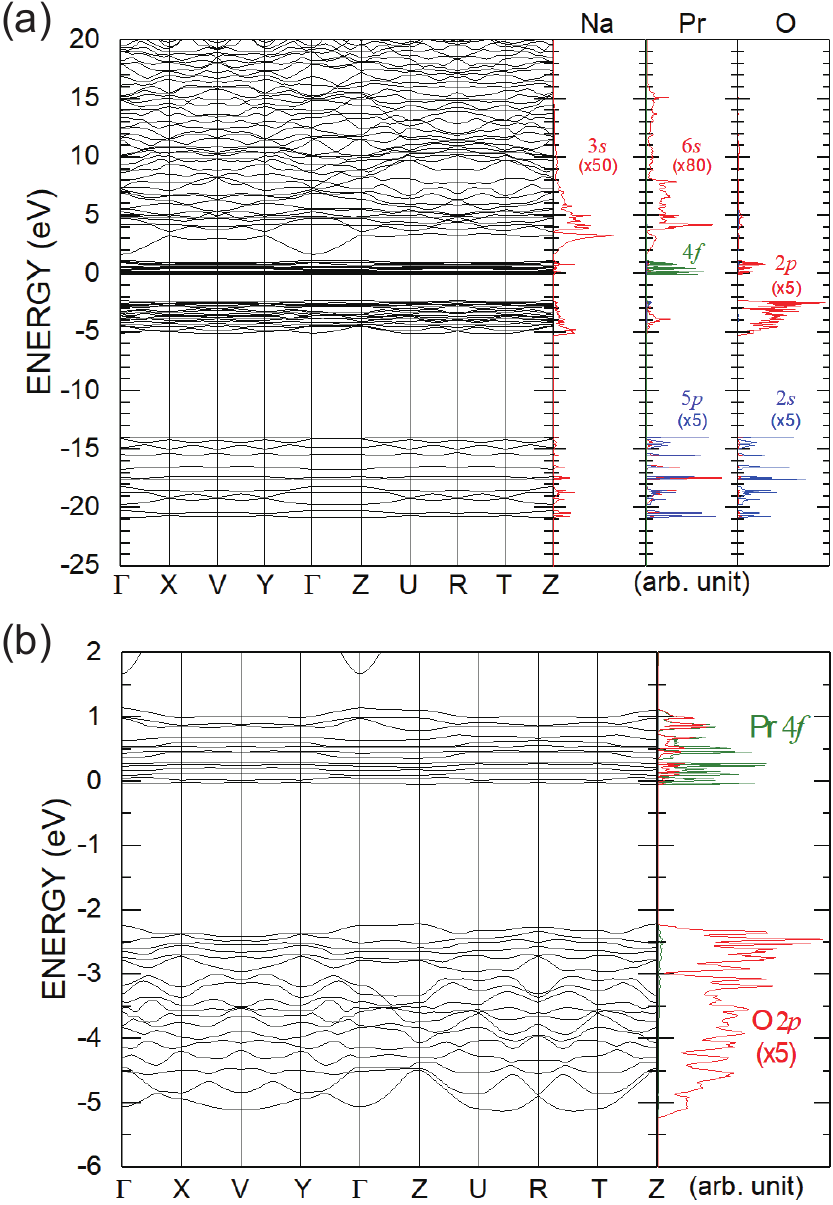}
\caption{\label{fig:fs2}Electronic band structures and the projected density of states for Na$_2$PrO$_3$ in the energy range from (a) $-25$~eV to $20$~eV and (b) $-6$~eV to $2$~eV. The Fermi level is set to zero.}
\end{figure}

We construct the MLWFs for the Pr $4f$ and O $2p$ bands in Figs.~\ref{fig:fs1}(b) and \ref{fig:fs2}(b). From the MLWFs, we estimate the transfer integrals for constructing the tight-binding model. We here present the values for the Pr $4f$ orbitals at the nearest-neighbor sites on a $z$ bond for Li$_2$PrO$_3$ and Na$_2$PrO$_3$ in Tables~\ref{tab:table_s1} and \ref{tab:table_s2}, respectively. The transfer integrals are calculated as $t_{\mu\nu} = \langle i,\mu| \mathcal{H}_0 |j,\nu \rangle$, where $\mathcal{H}_0$ is the {\it ab initio} Hamiltonian, $|i,\nu\rangle$ is the MLWF at site $i$ with orbital $\nu$($=\xi$, $\eta$, $\zeta$, $A$, $\alpha$, $\beta$, and $\gamma$); $i$ and $j$ denote the neighboring sites on a $z$ bond. The values include both direct and indirect (via O $2p$) paths, and averaged over three types of bonds to recover the C$_3$ symmetry. For simplicity, we focus on the spin diagonal components; the off-diagonal ones mixing different spins are small (the absolute values are all less than 1~meV), and neglected in the perturbation in the main text. These values as well as the transfer integrals between the neighboring Pr $4f$ and O $2p$ orbitals are used for the tight-binding band structures in Fig.~3 in the main text. As shown in Tables~\ref{tab:table_s1} and \ref{tab:table_s2}, the most dominant transfer integrals are $t_{\xi\alpha}(=-t_{\eta\beta}^{*})$ and $t_{\zeta\zeta}$, both of which are discussed in the main text to give dominant contributions to the antiferromagnetic Kitaev coupling $K$; see Fig.~5 in the main text. Note that the diagonal components for $\alpha$ and $\beta$ are large but do not contribute to the perturbation.

\newpage

\begin{table}[t]
\centering
\caption{\label{tab:table_s1}The nearest-neighbor transfer integrals $t_{\mu\nu}$ on a $z$ bond for Li$_2$PrO$_3$; $\mu$ is in the row and $\nu$ is in the column. The unit is in meV. The upper-right half of the table is omitted as the matrix is Hermite conjugate. See the text for details.}
\begin{ruledtabular}
\begin{tabular}{cccccccc}
&$\xi$&$\eta$&$\zeta$&$A$&$\alpha$&$\beta$&$\gamma$\\
\hline
$\xi$&$32.4$& & & & & & \\
$\eta$&$29.8+0.4i$&$32.4$& & & & & \\
$\zeta$&$3.51+0.04i$&$3.51-0.04i$&$-75.9$& & & & \\
$A$&$-2.14-0.17i$&$2.14-0.17i$&$0.01-0.55i$&$-57.1$& & & \\
$\alpha$&$-87.9+0.8i$&$1.13+0.51i$&$-9.22+0.12i$&$4.07+0.55i$&$153$& & \\
$\beta$&$-1.13+0.51i$&$87.9+0.8i$&$9.22+0.12i$&$4.07-0.55i$&$-23.2- 1.7i$&$153$& \\
$\gamma$&$-6.17+0.33i$&$6.17+0.33i$&$-0.03-0.40i$&$-35.1$&$-0.28+0.42i$&$-0.28-0.42i$&$40.8$\\ 
\end{tabular}
\end{ruledtabular}
\end{table}

\begin{table}[t]
\centering
\caption{\label{tab:table_s2}The nearest-neighbor transfer integrals $t_{\mu\nu}$ on a $z$ bond for Na$_2$PrO$_3$; $\mu$ is in the row and $\nu$ is in the column. The unit is in meV. The upper-right half of the table is omitted as the matrix is Hermite conjugate. See the text for details.}
\begin{ruledtabular}
\begin{tabular}{cccccccc}
&$\xi$&$\eta$&$\zeta$&$A$&$\alpha$&$\beta$&$\gamma$\\
\hline
$\xi$&$14.6$& & & & & & \\
$\eta$&$-4.47+0.22i$&$14.6$& & & & & \\
$\zeta$&$-0.61-0.01i$&$-0.61+0.01i$&$-84.9$& & & & \\
$A$&$2.86-0.11i$&$-2.86-0.11i$&$0.03-0.62i$&$-34.4$& & & \\
$\alpha$&$-65.5+0.2i$&$12.7+0.3i$&$8.88-0.21i$&$-6.65+0.33i$&$132$& & \\
$\beta$&$-12.7+0.3i$&$65.5+0.2i$&$-8.88-0.21i$&$-6.65-0.33i$&$-47.7- 1.3i$&$132$& \\
$\gamma$&$0.55+0.47i$&$-0.55+0.47i$&$-0.02-0.23i$&$-11.5$&$4.33+0.41i$&$4.33-0.41i$&$45.0$\\ 
\end{tabular}
\end{ruledtabular}
\end{table}

\end{document}